\documentclass[aps,twocolumn,showpacs,preprintnumbers,prb]{revtex4}
\usepackage{graphicx}

\newcommand{\be}{\begin{equation}}
\newcommand{\ee}{\end{equation}}
\newcommand{\bea}{\begin{eqnarray}}
\newcommand{\eea}{\end{eqnarray}}
\newcommand{\bean}{\begin{eqnarray*}}
\newcommand{\eean}{\end{eqnarray*}}

\begin{document}

\title{Wavevector analysis of the jellium exchange-correlation surface
energy in the random-phase approximation: detailed support for nonempirical
density functionals}
\author{ J. M. Pitarke$^{1,2}$, Lucian A. Constantin$^3$, and John P.
Perdew$^3$}
\affiliation{$^1$Materia Kondentsatuaren Fisika Saila, Zientzi
Fakultatea, Euskal Herriko Unibertsitatea\\
644 Posta kutxatila, E-48080 Bilbo, Basque Country\\
$^2$Donostia International Physics Center (DIPC) and Unidad F\'\i sica
Materiales
CSIC-UPV/EHU,\\
Manuel de Lardizabal Pasealekua, E-20018 Donostia, Basque Country\\
$^3$Department of Physics and Quantum Theory Group, Tulane University, New
Orleans, LA 70118}

\date{\today}

\begin{abstract}
We report the first three-dimensional wavevector analysis of the jellium
exchange-correlation (xc) surface energy in the random-phase approximation
(RPA). The RPA accurately describes long-range xc effects which are
challenging for semi-local approximations, since it includes the
universal small-wavevector behavior derived by Langreth and Perdew. We use
these rigorous RPA calculations for jellium slabs to test 
RPA versions of nonempirical semi-local
density-functional approximations for the xc energy. The local spin density
approximation (LSDA) displays cancelling errors in the small and intermediate
wavevector regions. The PBE GGA improves the analysis for intermediate
wavevectors, but remains too low for small wavevectors (implying too-low
jellium xc surface energies). The nonempirical meta-generalized gradient
approximation of Tao, Perdew, Staroverov, and Scuseria (TPSS meta-GGA) gives a
realistic wavevector analysis, even for small wavevectors or long-range
effects. We also study the effects of slab thickness and of short-range 
corrections to RPA.
\end{abstract}

\pacs{71.10.Ca,71.15.Mb,71.45.Gm}

\maketitle

\section{Introduction}
\label{sec1}

Modern electronic-structure calculations for atoms, molecules, and solids
usually rely upon Kohn-Sham (KS) density-functional theory (DFT),\cite{Ab,Ab2}
in which only $E_{xc}[n]$, the exchange-correlation (xc) energy as a
functional of electron density, must be approximated. Semi-empirical
approximations tend to be limited to systems that resemble those in the fitted
data set (typically small molecules), but nonempirical ones are constructed to
satisfy universal constraints and so should have a wider range of
applicability.\cite{Bb} For example, it is expected that a good description
of chemical reactions at a solid surface requires a good description of both
the molecules and the surface.

Jellium is a simple model of a simple metal, in which the valence electrons are
neutralized by a uniform positive background that extends up to a sharp planar
surface. The apparent success of the simplest density functional, the local
spin density approximation (LSDA), for the jellium surface energy\cite{LK}
motivated early interest in density functionals and in refinements of the LSDA
such as the generalized gradient approximation (GGA).\cite{LP1,Cb}

It was therefore a matter of some concern when wavefunction-based Fermi
HyperNetted-Chain (FHNC)\cite{fhnc} and fixed-node Diffusion Monte Carlo
(DMC)\cite{Db} calculations for jellium slabs (and their extrapolation to
infinite thickness) predicted surface energies considerably higher than those
obtained in the LSDA. Indeed, DMC is usually a gold standard of accuracy.
However, it encounters special difficulties for jellium slabs;\cite{Eb}
furthermore, the large deviations between the available DMC and LSDA
calculations have been attributed in part to inconsistency between the energy
of the inhomogeneous system and that of the corresponding homogeneous electron
gas.\cite{PP,PO4} Recent approaches\cite{PP,PO4,PE2,YPKFA,APF,JGDG,CPT} have
all suggested that the actual jellium surface energies are only a little
higher than those obtained in the LSDA. The jellium surface-energy story is
presented in full detail in Ref.~\onlinecite{CPT}.

In this paper, we perform a detailed analysis of exchange and correlation in
jellium slabs, {\it exact} at the level of the random phase approximation
(RPA), to show that the most refined nonempirical density functional, the
meta-generalized gradient approximation of Tao, Perdew, Staroverov, and
Scuseria (TPSS meta-GGA),\cite{TPSS1} can account even for the most
long-ranged xc effects at a jellium surface. This is a considerable
achievement for a semi-local functional that is inherently more reliable for
short-ranged effects than for long-ranged ones. RPA is known to be correct at
long range; because it has serious deficiencies at short-range and, therefore,
cannot be compared to standard versions of the semi-local functionals, we
use RPA versions of these functionals in this test.

In order to separate long-range and short-range xc effects, we look at the
surface contribution to the spherically-averaged real-space xc hole, averaged
over the electron density of the system, and its Fourier transform (wavevector
analysis). Langreth and Perdew\cite{LP1} showed that the exact xc energy of an
arbitrary inhomogeneous system can be obtained from a three-dimensional (3D)
Fourier transform of the spherical average of the xc hole density, which is a
function of a 3D wavevector ${\bf k}$. In the case of a plane-bounded electron
gas, this wavevector-dependent spherical average is dominated at long
wavelengths ($k\to 0$) by the zero-point energy-shift of the newly created
surface collective oscillations (surface plasmons) and takes a simple
analytical form. This known limit has been used to carry out a wavevector
interpolation correction to LSDA,\cite{LP1} PBE-GGA,\cite{YPKFA} and
TPSS-metaGGA\cite{CPT} xc surface energies.
The wavevector interpolation corrections to these functionals were
controlled\cite{YPKFA,CPT} by using the exact RPA values reported in
Ref.~\onlinecite{PE2}, and led to a consistent set of predicted surface
energies.\cite{CPT}

In a DFT context, the RPA is based upon the time-dependent Hartree
approximation for the density-response function but replacing the occupied and
unoccupied single-particle Hartree orbitals and energies by the corresponding
eigenfunctions and eigenvalues of the KS Hamiltonian of DFT.\cite{LP1} Hence,
it
describes the exchange energy and the long-range part of the correlation
energy correctly. Essentially {\it exact} RPA surface energies were evaluated
from
single-particle LSDA orbitals and energies in Ref.~\onlinecite{PE2}. These
calculations provide an accurate standard against which approximate density
functionals (in their RPA versions) can be tested and normed. The RPA versions
of LSD and GGA were reported in Refs.~\onlinecite{PW1} and~\onlinecite{YPK},
respectively. Because RPA is not self-correlation-free, the GGA for RPA
correlation is its own meta-GGA. The RPA version of the
nonempirical TPSS meta-GGA was investigated in Ref.~\onlinecite{CPT}.

Unless stated otherwise, atomic units are used throughout, i.e.,
$e^2=\hbar=m_e=1$. 

\section{Theoretical Framework}
\label{sec2}

The exact xc energy, $E_{xc}[n]$, of an arbitrary inhomogeneous system of
density $n({\bf r})$ can be obtained from the spherical average
$\bar n_{xc}({\bf r},u)$ of the coupling-constant averaged xc hole density
$\bar n_{xc}({\bf r},{\bf r}')$ at ${\bf r}'$ around an electron at ${\bf r}$,
as
follows\cite{LP1,CPT}
\begin{equation}
E_{xc}[n]=\int d{\bf r}\,n({\bf r})\,\varepsilon_{xc}[n]({\bf r}),
\end{equation}
where $\varepsilon[n]({\bf r})$ represents the xc energy per particle at point
${\bf r}$:
\begin{equation}\label{new}
\varepsilon_{xc}[n]({\bf r})=4\int^{\infty}_{0} dk\int^{\infty}_{0} du\,u^2\,
{\sin ku\over ku}\,\bar n_{xc}({\bf r},u),
\end{equation}
with
\begin{equation}\label{average}
\bar n_{xc}({\bf r},u)={1\over 4\pi}\int d\Omega\,\bar n_{xc}({\bf r},{\bf
r}'),
\end{equation}
$d\Omega$ being a differential solid angle around the direction of
${\bf u}={\bf r}'-{\bf r}$.

The xc surface energy, $\sigma_{xc}$, is obtained by subtracting from the xc
energy $E_{xc}[n]$ of a semi-infinite electron system the corresponding energy
$E_{xc}^\mathrm{unif}(n)$ of a uniform electron gas. In a jellium model, in
which the
electron system is translationally invariant in the plane of the surface, and
assuming the surface to be normal to the $z$-axis, one finds
\begin{equation}
\sigma_{xc}=\int^{\infty}_{0}d\left(k\over 2k_F\right)\,\gamma_{xc}(k),
\label{va2}
\end{equation}
where\cite{notem1}
\begin{equation}\label{gamma}
\gamma_{xc}(k)=2\,{k_F\over\pi}\int_{-\infty}^{+\infty}dz\,n(z)\,b_{xc}(k,z),
\end{equation}
with $k_F=(3\pi^2\bar n)^{1/3}$, $\bar n$ being the background density, and
\begin{equation}
b_{xc}(k,z)=4\pi\int_0^\infty du\,u^2\,
{\sin ku\over ku}\left[\bar n_{xc}(z,u)-\bar n^\mathrm{unif}_{xc}(u)\right].
\label{va3}
\end{equation}
Alternatively, one can introduce Eq.~(\ref{average}) into Eq.~(\ref{va3}) to
find:
\begin{eqnarray}
b_{xc}(k,z)&=&
{1\over 2}\int_{-k}^{+k}{dk_{z}\over k}\int_{-\infty}^{+\infty}dz'\,
{\mathrm e}^{ik_z(z-z')}\cr\cr
&\times&\bar n_{xc}(k_{\parallel};z,z')-\bar n_{xc}^\mathrm{unif}(k),
\label{wvj6}
\end{eqnarray}
with $k_\parallel=\sqrt{k^2-k_z^2}$, and $\bar n_{xc}(k_{\parallel};z,z')$ and
$\bar n_{xc}^\mathrm{unif}(k)$ representing Fourier transforms of the
coupling-constant averaged xc hole densities $\bar n_{xc}({\bf r},{\bf r}')$
and $\bar n_{xc}^\mathrm{unif}({\bf r},{\bf r}')$, respectively. At long
wavelengths
($k\to 0$), one finds the exact limit\cite{LP1}
\begin{equation}\label{lp} 
\gamma_{xc}(k)={k_F\over 4\pi}\left(\omega_s-{1\over 2}\omega_p\right)k,
\label{xcs1}
\end{equation}
which only depends on the bulk- and surface-plasmon energies
$\omega_p=(4\pi\bar n)^{1/2}$ and $\omega_s=\omega_p/\sqrt 2$,
and does not depend, therefore, on the electron-density profile at the surface.

The spherical average $\bar n_{xc}(z,u)$ entering Eq.~(\ref{va3}) can be
obtained within local or semi-local density-functional approximations (such as
LSDA, PBE GGA, and TPSS meta-GGA) from models \cite{pw2,YPK,a113,CPT} that
require
knowledge of the xc hole density
$\bar{n}_{xc}^\mathrm{unif}(u)$ of a uniform electron gas. Alternatively,
rigorous
calculations of
$\bar n_{xc}^\mathrm{unif}(k)$ and the fully nonlocal
$\bar n_{xc}(k_\parallel;z,z')$ entering Eq.~(\ref{wvj6}) can be carried out
from knowledge of the $\lambda$-dependent density-response functions
$\chi_\mathrm{unif}^\lambda(k,\omega)$ and $\chi^\lambda(k_\parallel
\omega;z,z')$,
respectively, defined by adiabatically switching on the e-e interaction via the
coupling constant $\lambda$ and by adding, at the same time, an external
potential so as to maintain the true ($\lambda=1$) ground-state density in the
presence of the modified e-e interaction.\cite{a30,gl} By using the
fluctuation-dissipation theorem,\cite{a109,pines} one finds:
\begin{equation}
\bar n_{xc}^\mathrm{unif}(k)={1\over\bar n}
\left[-\frac{1}{\pi}\int_0^1d\lambda\int_0^\infty
d\omega\chi_\mathrm{unif}^\lambda(k,i\omega)-\bar n\right]
\label{wvj5}
\end{equation}
and
\begin{eqnarray}
\bar n_{xc}(k_{\parallel};z,z')=&-&{1\over\pi n(z)}\int^{1}_{0}
d\lambda\int^{\infty}_{0}d\omega
\,\chi^\lambda(k_{\parallel},i\omega;z,z')\cr\cr&-&\delta(z-z').
\label{wvj7}
\end{eqnarray}

With the aim of testing the performance of local and semi-local
density-functional approximations for the xc surface energy, we compare these
(local and semi-local) calculations [obtained from Eq.~(\ref{va3})] to their
fully nonlocal counterparts [obtained from Eq.~(\ref{wvj6}) with the aid of
Eqs.~(\ref{wvj5}) and (\ref{wvj7})] at the same level of approximation, which
we choose to be the RPA. On the one hand, we evaluate $\gamma_{xc}(k)$ from RPA
versions (LSDA-RPA, PBE-RPA, and TPSS-RPA) of the local (or semi-local) $\bar
n_{xc}(z,u)$ entering Eq.~(\ref{va3}) based on the RPA xc hole density $\bar
n_{xc}^\mathrm{unif}(u)$ of a uniform electron gas. On the other hand, we
evaluate
$\gamma_{xc}(k)$ from a fully nonlocal version (exact-RPA) of
$\bar n_{xc}(k_\parallel;z,z')$ entering Eq.~(\ref{wvj6}) based [by
using Eq.~(\ref{wvj7})] on the RPA density-response
function $\chi^\lambda(k_\parallel,\omega;z,z')$. 

\section{Results}
\label{sec3}

In the calculations presented below, we have considered a jellium slab of
background thickness $a=2.23\,\lambda_F$, $\lambda_F$ being the Fermi
wavelength ($\lambda_F=2\pi/k_F$), and background density
$\bar n=[(4\pi/3)r_s^3]^{-1}$ with $r_s=2.07$. This slab
corresponds to about
four atomic layers of Al(100).

For the LSDA-RPA calculations, we have obtained the RPA xc hole density
$\bar n_{xc}^\mathrm{unif}(u)$ of a uniform electron gas either from
Eq.~(\ref{wvj5})
or from a non-oscillatory parametrization.\cite{notex} For the PBE-RPA and
TPSS-RPA calculations, we have always used a non-oscillatory parametrization
of the RPA xc hole density $\bar n_{xc}^\mathrm{unif}(u)$.\cite{notex}

For the evaluation of the fully nonlocal (exact-RPA) $\gamma_{xc}(k)$ of
Eq.~(\ref{gamma}), we follow the method described in Ref.~\onlinecite{PE2}. We
first assume that $n(z)$ vanishes at a distance $z_0$ from either jellium
edge,\cite{note1} and we expand the single-particle wave functions $\phi_l(z)$
and the density-response function $\chi^\lambda(k_\parallel,\omega;z,z')$ in
sine and double-cosine Fourier representations, respectively. We then perform
the integrals over the coordinates $z$ and $z'$ analytically, and we find an
explicit expression for $\gamma_{xc}(k)$ [see Eqs.~(\ref{wvj14})-(\ref{last})
of the Appendix] in terms of the single-particle energies $\varepsilon_l$ and
the Fourier coefficients $b_{ls}$ and $\chi_{mn}(k_\parallel,\omega)$ of the
single-particle wave functions $\phi_l(z)$ and the density-response function
$\chi^\lambda(k_\parallel,\omega;z,z')$, respectively.\cite{eguiluz} We have
taken all the single-particle wave functions $\phi_l(z)$ and energies
$\varepsilon_l$ to be the LDA eigenfunctions and eigenvalues of the KS
Hamiltonian of DFT, as obtained by using the Perdew-Wang
parametrization\cite{PW1} of the Ceperley-Alder xc energy of the homogeneous
electron gas.\cite{ca} For the jellium slab with $r_s=2.07$ and
$a=2.23\lambda_F$ considered here, the {\it exact} RPA xc surface energy is
found to be $\sigma_{xc}=3091\,\mathrm{erg}/\mathrm{cm}^2$, not far from the
corresponding RPA xc surface energy of a semi-infinite jellium which is known
to be $\sigma_{xc}=3064\,\mathrm{erg}/\mathrm{cm}^2$.\cite{PP}

%%%%%%%%%%%%%%%%%%%%%%%%%%%%%%%%%%%%%%%%%%%%%%%%%%%%%
\begin{figure}
\includegraphics[width=\columnwidth]{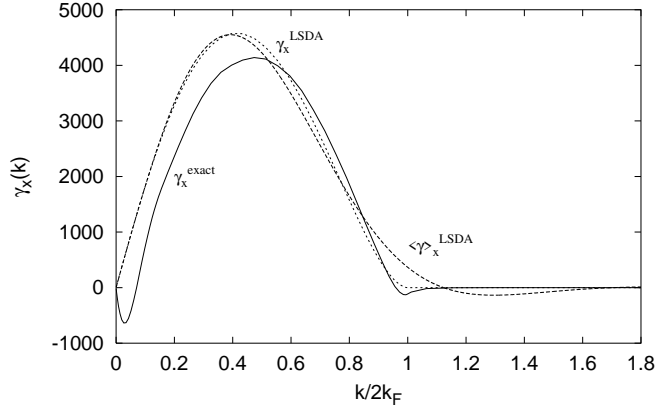}
\caption{Wavevector analysis $\gamma_x(k)$, versus $k/2k_F$, of the exchange
surface energy of a jellium slab of thickness $a=2.23\lambda_F$ and
$r_s=2.07$. Solid and dashed lines represent {\it exact} and LSDA
calculations, respectively. The LSDA calculation has been performed either
from the actual exchange hole density $n_x^\mathrm{unif}(u)$ of a uniform
electron gas, which we have obtained from Eq.~(\ref{wvj5}) with
$\chi_\mathrm{unif}^\lambda(k,\omega)$ replaced by $\chi_\mathrm{unif}^0(k
\omega)$, ($\gamma_x^\mathrm{LSDA}$) or from the non-oscillatory
parametrization of $n_x^\mathrm{unif}(u)$ reported in Ref.~\onlinecite{a113}
($<\gamma>_x^\mathrm{LSDA}$). The area under each curve represents the exchange
surface energy: $\sigma_x^\mathrm{LSDA}=2699\,\mathrm{erg}/\mathrm{cm}^2$ and
$\sigma_x^\mathrm{exact}=2348\,\mathrm{erg}/\mathrm{cm}^2$. 
(1 $\mathrm{hartree}/\mathrm{bohr}^{2} = 1.557\times 10^{6}\,
\mathrm{erg}/\mathrm{cm}^{2}.$)}
\label{wvjf5}
\end{figure}
%%%%%%%%%%%%%%%%%%%%%%%%%%%%%%%%%%%%%%%%%%%%%%%%%%%%%%%

%%%%%%%%%%%%%%%%%%%%%%%%%%%%%%%%%%%%%%%%%%%%%%%%%%%%%
\begin{figure}
\includegraphics[width=\columnwidth]{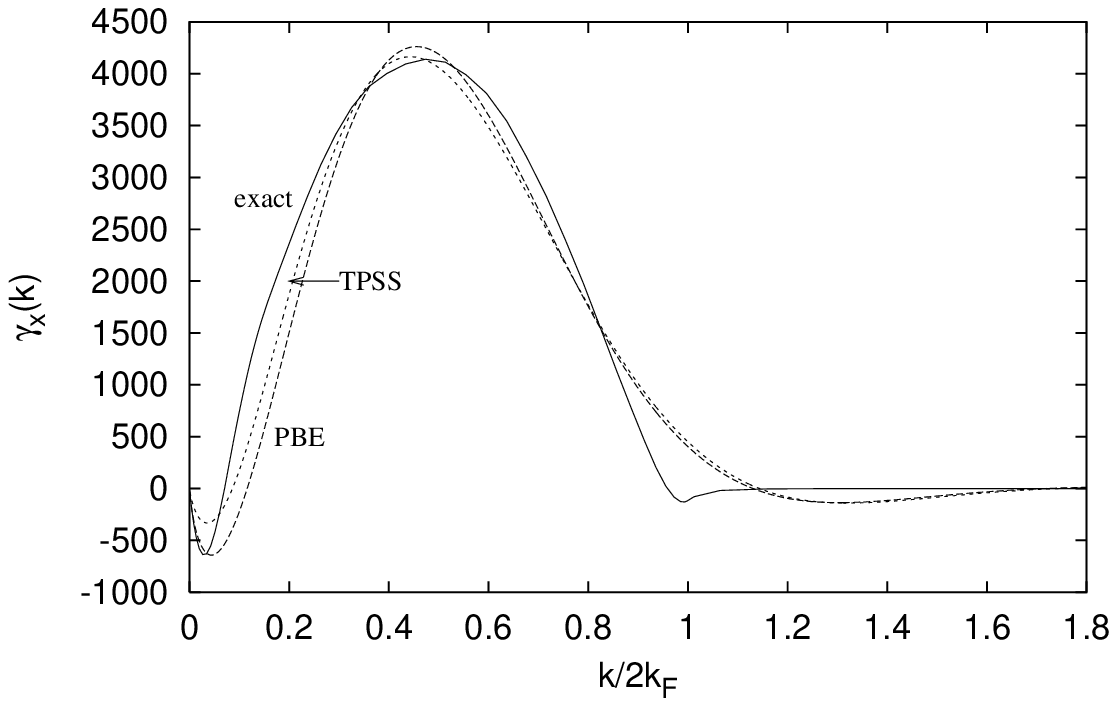}
\caption{Wavevector analysis $\gamma_x(k)$, versus $k/2k_F$, of the exchange
surface energy of a jellium slab of thickness $a=2.23\lambda_F$ and
$r_s=2.07$. Solid and dashed lines represent {\it exact} and semi-local (PBE
and TPSS) calculations, respectively. The semi-local PBE and TPSS calculations
have been performed from the non-oscillatory parametrization of
$n_x^\mathrm{unif}(u)$ reported in Ref.~\onlinecite{a113}. The area under each
curve represents the exchange surface energy:
$\sigma_x^\mathrm{PBE}=2155\,\mathrm{erg}/\mathrm{cm}^2$,
$\sigma_x^\mathrm{TPSS}=2247\,\mathrm{erg}/\mathrm{cm}^2$, and
$\sigma_x^\mathrm{exact}=2348\,\mathrm{erg}/\mathrm{cm}^2$.}
\label{wvjf6}
\end{figure}
%%%%%%%%%%%%%%%%%%%%%%%%%%%%%%%%%%%%%%%%%%%%%%%%%%%%%%%

In Figs.~\ref{wvjf5} and~\ref{wvjf6} we have plotted (solid lines) the
exact-exchange contribution to $\gamma_{xc}(k)$, i.e., $\gamma_x(k)$, which we
have obtained from Eqs.~(\ref{wvj14})-(\ref{last}) with the quantities
$\chi_\mathrm{unif}^\lambda(k,\omega)$ and $\chi_{mn}^\lambda(k_\parallel
\omega)$
replaced by their noninteracting counterparts
$\chi_\mathrm{unif}^0(k,\omega)$ and $\chi_{mn}^0(k_\parallel,\omega)$,
respectively.
Also plotted in these figures are the LSDA, PBE, and TPSS
calculations of $\gamma_x(k)$ that we have obtained by replacing the xc hole
densities $\bar{n}_{xc}(z,u)$ and $\bar{n}_{xc}^\mathrm{unif}(u)$ entering
Eq.~(\ref{va3}) by their corresponding exchange-only counterparts (dashed
lines).

The LSDA $\gamma_x(k)$ represented in Fig.~\ref{wvjf5} has been obtained by
using both the actual exchange hole density $n_x^\mathrm{unif}(u)$ of a uniform
electron gas [dashed curve labeled $\gamma_x^\mathrm{LSDA}$], which we have
obtained
from Eq.~(\ref{wvj5}) with $\chi_\mathrm{unif}^\lambda(k,\omega)$ replaced by
$\chi_\mathrm{unif}^0(k,\omega)$, and the non-oscillatory
exchange hole density $n_x^\mathrm{unif}(u)$ reported in Ref.~\onlinecite{a113}
[dashed curve labeled $<\gamma>_x^\mathrm{LSDA}$].
$\gamma^\mathrm{LSDA}_{x}(k)$ and
$<\gamma>^\mathrm{LSDA}_{x}(k)$ yield, by construction of the non-oscillatory
exchange hole density $n_x^\mathrm{unif}(u)$, the same
exchange surface energy $\sigma_x$; they are also almost identical in a wide
range of low wavevectors, but $<\gamma>^\mathrm{LSDA}_{x}(k)$ is considerably
less
accurate near $k=2k_F$ where the exact $\gamma_x(k)$ has a kink. This kink is
realistic for jellium-like systems, but not for atoms and molecules. 

The PBE and TPSS $\gamma_x(k)$ represented in Fig.~\ref{wvjf6} have both been
obtained by using the non-oscillatory exchange hole density
$n_x^\mathrm{unif}(u)$ reported in Ref.~\onlinecite{a113}, which yields a wrong
behavior of $\gamma_x(k)$ at
large wavevectors. Nevertheless, both the actual exchange hole density
$n_x^\mathrm{unif}(u)$ of a uniform electron gas (not used in these
calculations) and
the corresponding non-oscillatory exchange hole density would yield the same
exchange surface energy $\sigma_x$, by construction, as occurs in the LSDA.

Figs.~\ref{wvjf5} and~\ref{wvjf6} show that while the LSDA $\gamma_x(k)$
considerably overestimates the exact $\gamma_x(k)$ at low wavevectors (see
Fig.~\ref{wvjf5}), leading to an exchange surface energy $\sigma_x$ that is too
large, the PBE and TPSS $\gamma_x(k)$ are close to the exact $\gamma_x(k)$ (see
Fig.~\ref{wvjf6}). We note that the peaks of $\gamma^\mathrm{PBE}_{x}(k)$ and
$\gamma^\mathrm{TPSS}_{x}(k)$ are close to the exact one, a fact which was used
in the
construction of the TPSS exchange hole,\cite{CPT} and that at larger
wavevectors $\gamma^\mathrm{PBE}_{x}(k)$ and $\gamma^\mathrm{TPSS}_{x}(k)$
nearly coincide, as expected; at lower wavevectors, however, the TPSS meta-GGA
differs from the
PBE GGA, leading to a wavevector-dependent $\gamma_x(k)$ that is closer to the
exact behavior. 

We have also carried out calculations of the exact $\gamma_x(k)$ for increasing
values of the background thickness $a$, and we have found that (i)
$\gamma_x(k)$ is only sensitive to the size of the system at wavevectors below
the minimum that is present in the solid lines of Figs.~\ref{wvjf5} and
\ref{wvjf6}, and (ii) as $k\to 0$ the wavevector-dependent $\gamma_x(k)$
approaches 
in the semi-infinite limit the profile-independent negative value
($\gamma_x=-1.50\times 10^4/r_s^3\,\mathrm{erg}/\mathrm{cm}^2$)
reported in Refs.~\onlinecite{LP1} and \onlinecite{rg}.

%%%%%%%%%%%%%%%%%%%%%%%%%%%%%%%%%%%%%%%%%%%%%%%%%%%%%
\begin{figure}
\includegraphics[width=\columnwidth]{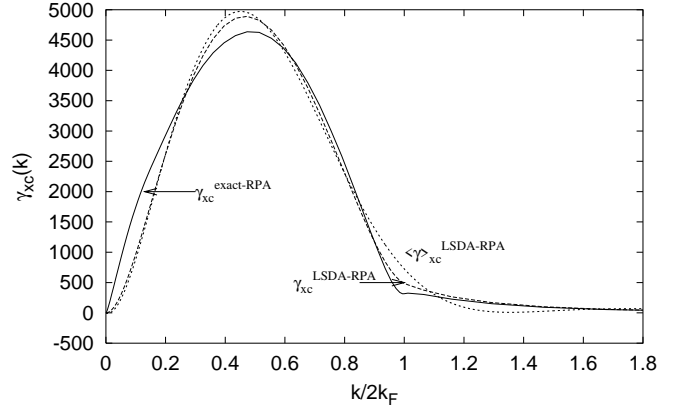}
\caption{Wavevector analysis $\gamma_{xc}(k)$, versus $k/2k_F$, of the RPA xc
surface energy of a jellium slab of thickness $a=2.23\lambda_F$ and
$r_s=2.07$. Solid and dashed lines represent exact-RPA and LSDA-RPA
calculations, respectively. The LSDA calculation has been performed either
from the actual RPA xc hole density of Eq.~(\ref{wvj5})
($\gamma_{xc}^\mathrm{LSDA-RPA}$) or from a non-oscillatory parametrization of
$\bar n_{xc}^\mathrm{unif}(u)$ ($<\gamma>_{xc}^\mathrm{LSDA}$).\cite{notex}
The area under each curve represents the RPA xc surface energy:
$\sigma_{xc}^\mathrm{LSDA-RPA}=3034\,\mathrm{erg}/\mathrm{cm}^2$ and
$\sigma_{xc}^\mathrm{exact-RPA}=3091\,\mathrm{erg}/\mathrm{cm}^2$.}
\label{wvjf7}
\end{figure}
%%%%%%%%%%%%%%%%%%%%%%%%%%%%%%%%%%%%%%%%%%%%%%%%%%%%%%%

%%%%%%%%%%%%%%%%%%%%%%%%%%%%%%%%%%%%%%%%%%%%%%%%%%%%%
\begin{figure}
\includegraphics[width=\columnwidth]{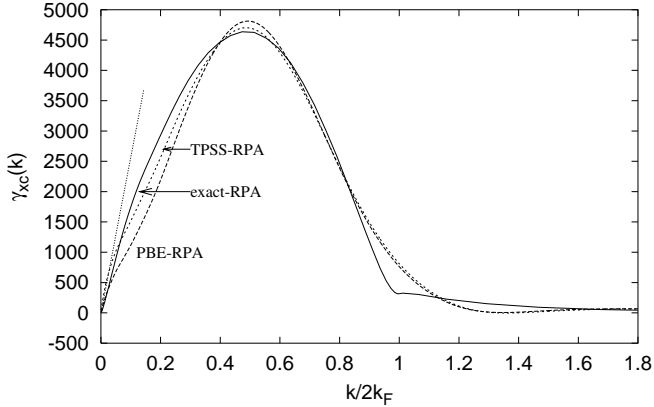}
\caption{Wavevector analysis $\gamma_{xc}(k)$, $k/2k_F$, of the RPA xc surface
energy of a jellium slab of thickness $a=2.23\lambda_F$ and $r_s=2.07$.
Solid and dashed lines represent exact-RPA and semilocal-RPA (PBE-RPA and
TPSS-RPA) calculations, respectively. The semilocal PBE-RPA and TPSS-RPA
calculations have been performed from a non-oscillatory parametrization of
$\bar n_{xc}^\mathrm{unif}(u)$.\cite{notex} The area under each curve
represents the RPA xc surface energy:
$\sigma_{xc}^\mathrm{PBE-RPA}=2959\,\mathrm{erg}/\mathrm{cm}^2$,
$\sigma_{xc}^\mathrm{TPSS-RPA}=3052\,\mathrm{erg}/\mathrm{cm}^2$, and
$\sigma_{xc}^\mathrm{exact-RPA}=3091\,\mathrm{erg}/\mathrm{cm}^2$. The straight
dotted line
represents the universal low-wavevector limit of Eq. (\ref{xcs1}).}
\label{wvjf8}
\end{figure}
%%%%%%%%%%%%%%%%%%%%%%%%%%%%%%%%%%%%%%%%%%%%%%%%%%%%%%%

%%%%%%%%%%%%%%%%%%%%%%%%%%%%%%%%%%%%%%%%%%%%%%%%%%%%%
\begin{figure}
\includegraphics[width=\columnwidth]{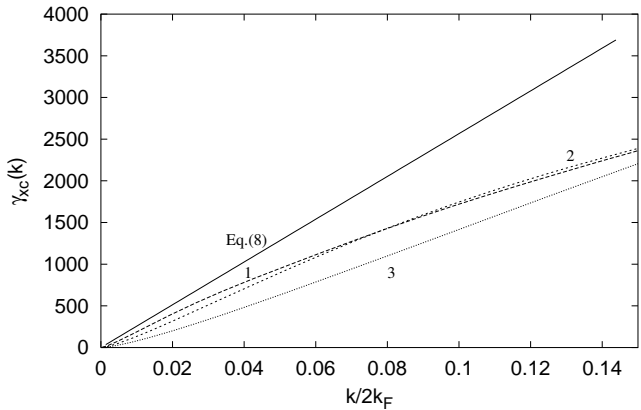}
\caption{Wavevector analysis $\gamma_{xc}(k)$, versus $k/2k_F$, of the {\it
exact} RPA xc surface energy of jellium slabs of $r_s=2.07$ and various values
of the background thickness: $a=8.23\lambda_{F}$ ('1'), $a=2.23\lambda_{F}$
('2'), and $a=0.56\lambda_{F}$ ('3'). The straight solid line represents the
universal low-wavevector limit of Eq.~(\ref{xcs1}), which corresponds to a
plane-bonded semi-infinite system ($a\to\infty$).}
\label{wvjf12}
\end{figure}
%%%%%%%%%%%%%%%%%%%%%%%%%%%%%%%%%%%%%%%%%%%%%%%%%%%%%%%

%%%%%%%%%%%%%%%%%%%%%%%%%%%%%%%%%%%%%%%%%%%%%%%%%%%%%
\begin{figure}
\includegraphics[width=\columnwidth]{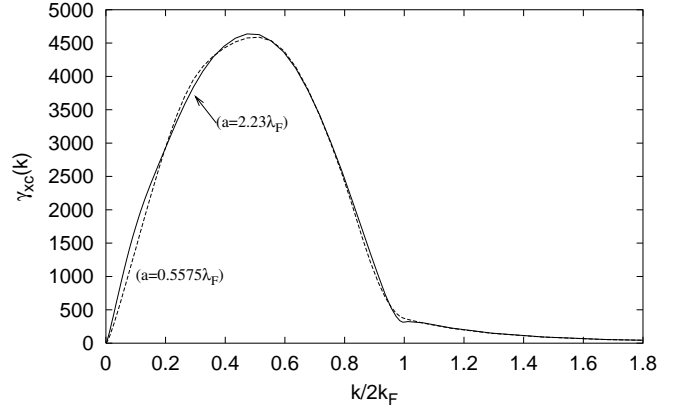}
\caption{Wavevector analysis $\gamma_{xc}(k)$, versus $k/2k_F$, of the
{\it exact} RPA xc surface energy of jellium slabs of $r_s=2.07$ and 
two values of the
background thickness: $a=2.23\lambda_{F}$ (solid line) and
$a=0.56\lambda_{F}$ (dashed line). The area under each curve represents the
exact RPA xc surface energy $\sigma_{xc}^\mathrm{exact-RPA}$:
$3091\,\mathrm{erg}/\mathrm{cm}^2$ and $3043\,\mathrm{erg}/\mathrm{cm}^2$, for
$a=2.23\lambda_F$ and $a=0.56\lambda_F$, respectively.}
\label{wvjf9}
\end{figure}
%%%%%%%%%%%%%%%%%%%%%%%%%%%%%%%%%%%%%%%%%%%%%%%%%%%%%%%

%%%%%%%%%%%%%%%%%%%%%%%%%%%%%%%%%%%%%%%%%%%%%%%%%%%%%
\begin{figure}
\includegraphics[width=\columnwidth]{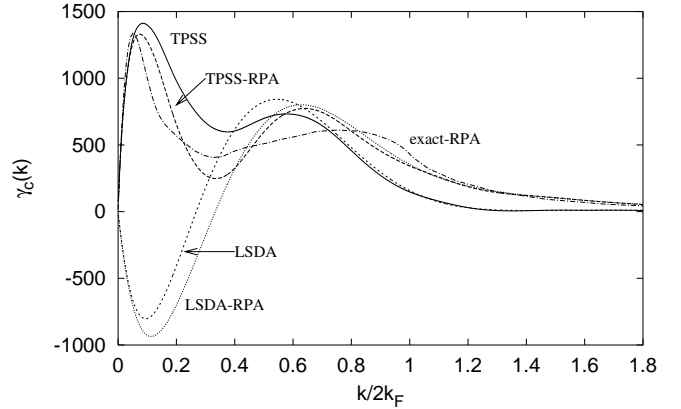}
\caption{Wavevector analysis $\gamma_{c}(k)$, versus $k/2k_F$, of the
correlation surface energy of a jellium slab of thickness $a=2.23\lambda_F$
and $r_s=2.07$. Dotted, long-dashed, and dashed-dotted lines represent
LSDA-RPA, TPSS-RPA, and exact-RPA calculations, respectively. Short-dashed and
solid lines represent {\it standard} versions of the LSDA and the semi-local
TPSS, as obtained from an accurate (beyond RPA) non-oscillatory
parametrization of the correlation hole density $\bar n_c^\mathrm{unif}(u)$ of
a uniform electron gas.\cite{pw2} The area under each curve represents the
correlation surface energy:
$\sigma_{c}^\mathrm{LSDA-RPA}=336\,\mathrm{erg}/\mathrm{cm}^2$,
$\sigma_{c}^\mathrm{TPSS-RPA}=804\,\mathrm{erg}/\mathrm{cm}^2$,
$\sigma_{c}^\mathrm{exact-RPA}=743\,\mathrm{erg}/\mathrm{cm}^2$,
$\sigma_{c}^\mathrm{LSDA}=290\,\mathrm{erg}/\mathrm{cm}^2$, and
$\sigma_{c}^\mathrm{TPSS}=756\,\mathrm{erg}/\mathrm{cm}^2$.}
\label{wvjf10}
\end{figure}
%%%%%%%%%%%%%%%%%%%%%%%%%%%%%%%%%%%%%%%%%%%%%%%%%%%%%%%   

Figures~\ref{wvjf7} and \ref{wvjf8} exhibit the results that we have obtained
for the RPA $\gamma_{xc}(k)$ from Eqs.~(\ref{wvj14})-(\ref{last}) (solid
lines) and within the LSDA-RPA, PBE-RPA, and TPSS-RPA (dashed lines). As in
the case of the exchange-only contributions represented in Figs.~\ref{wvjf5}
and~\ref{wvjf6}, the LSDA $\gamma_{xc}(k)$ represented in Fig.~\ref{wvjf7} has
been obtained by using both the actual RPA xc hole density
$\bar{n}_{xc}^\mathrm{unif}(u)$ [dashed line labeled $\gamma_{xc}^{LSDA-RPA}$],
which we have obtained from Eq.~(\ref{wvj5}), and
a non-oscillatory xc hole density $\bar{n}_{xc}^\mathrm{unif}(u)$ [dashed line
labeled $<\gamma>_{xc}^{LSDA-RPA}$]; the PBE and TPSS $\gamma_{xc}(k)$
represented in Fig.~\ref{wvjf8} have both been obtained by using a
non-oscillatory xc hole density $\bar{n}_{xc}^\mathrm{unif}(u)$.

Figure~\ref{wvjf7} shows that at short wavelengths with $k>2k_F$ the quantities
$\gamma_{xc}^{LSDA-RPA}(k)$ (dashed line) and $\gamma_{xc}^{exact-RPA}(k)$
(solid line) nearly coincide, as expected.\cite{LP1,a33,a71} The LSDA, however,
considerably underestimates $\gamma_{xc}(k)$ at low wavevectors.
This is {\it partially} compensated by an LSDA $\gamma_{xc}(k)$ that at
intermediate wavevectors (around the peak of $\gamma_{xc}(k)$) 
is too large. Figure~\ref{wvjf8} shows that the PBE GGA improves
$\gamma_{xc}(k)$ at intermediate wavevectors more than at low wavevectors,
thereby yielding a xc surface energy that is even smaller than in the LSDA.
From a different perspective,\cite{PCSB} the too-small PBE surface energy
arises from a too-large gradient coefficient for exchange, but this is
repaired by the TPSS meta-GGA which uses the proper gradient coefficient.
Indeed, Fig.~\ref{wvjf8} clearly shows that the TPSS meta-GGA brings
improvements over the corresponding PBE GGA at both intermediate and small
wavevectors, thus leading to a wavevector-dependent
$\gamma^{TPSS-RPA}_{xc}(k)$ that is very close to
$\gamma^{\mathrm{exact}-RPA}_{xc}(k)$ (solid line)
and to an xc surface energy $\sigma_{xc}$ that is only slightly lower than its
exact RPA counterpart.\cite{note2} We have obtained similar results (not
displayed here) for $r_s=3$, and we have found that the errors introduced by
the use of nonempirical semi-local density-functional approximations slightly
increase with $r_s$ as expected from the analysis of Ref.~\onlinecite{CPT}. 

Also represented in Fig.~\ref{wvjf8} (by a dotted line) is the universal
(density-profile independent) low-wavevector limit of Eq.~(\ref{lp}). The
TPSS-RPA $\gamma_{xc}(k)$ has the virtue that not only is it very close to its
exact-RPA counterpart in the whole range of low and intermediate wavevectors,
but it imitates the exact low-wavevector limit of Eq.~(\ref{lp}) as well.
That this limit is also reproduced by the exact-RPA $\gamma_{xc}(k)$ of a
semi-infinite electron system is shown in Fig.~\ref{wvjf12}, where we have
plotted calculations of this quantity for increasing values of the background
thickness $a$, from $a=0.56\lambda_F$ to $a=8.23\lambda_F$. Furthermore,
Fig.~\ref{wvjf9} shows that $\gamma_{xc}(k)$ is only sensitive to the
background thickness at very low wavevectors.

Finally, in order to investigate the impact of short-range corrections to the
RPA $\gamma_{xc}(k)$, we have plotted in Fig.~\ref{wvjf10} the correlation
contribution to $\gamma_{xc}(k)$, i.e, $\gamma_{c}(k)$, as obtained in the RPA
(LSDA-RPA, TPSS-RPA, and exact-RPA) and also in {\it standard} versions of
local and semi-local density-functionals (LSDA and TPSS) that use an accurate
(beyond RPA) non-oscillatory parametrization of the correlation hole density
$\bar n_c^\mathrm{unif}(u)$ of a uniform electron gas.\cite{pw2} We observe
that in
the
long-wavelength limit ($k\to 0$), where both LSDA-RPA and {\it standard} LSDA
exhibit serious deficiencies, both TPSS-RPA and the more accurate
{\it standard} TPSS coincide with the exact-RPA. At shorter wavelengths, the
{\it standard} TPSS predicts a substantial correction to its TPSS-RPA and
exact-RPA counterparts, which is first positive and then negative and leads,
therefore, to a persistent cancellation of short-range correlation effects
beyond the RPA similar to the cancellation that was reported in
Ref.~\onlinecite{PP} in the framework of time-dependent density-functional
theory and a two-dimensional wavevector analysis of the correlation surface
energy.     

\section{Conclusions}
\label{sec4}

We have reported the first 3D wavevector analysis of the jellium
xc surface energy in the RPA, and we have used this fully nonlocal (esentially
exact) RPA calculation to test RPA versions of nonempirical semi-local
density-functional approximations for the xc energy. We have tested the first
three-rungs of the Jacob's ladder classification of nonempirical density
functionals:\cite{PS} LSDA, PBE GGA, and TPSS meta-GGA.

We have found that while the LSDA displays cancelling errors in the small and
intermediate wavevector regions and the PBE GGA improves the analysis for
intermediates wavevectors while remaining too low for small wavevectors
(implying two-low xc surface energies), the TPSS meta-GGA yields a
realistic wavevector analysis even for small wavevectors or long-range effects.
We have also demonstrated numerically the correctness of the LSDA at large
wavevectors\cite{LP1,a33,a71} (where LSD-RPA, TPSS-RPA, and the exact-RPA
coincide, as shown in Fig.~\ref{wvjf10}) and the universal low-wavevector
behavior derived by Langreth and Perdew,\cite{LP1} which is nicely reproduced
by the TPSS meta-GGA.

We have carried out fully nonlocal RPA calculations for increasing values of
the background thickness, and we have found that the 3D wavevector analysis of
the xc surface energy is remarkably insensitive to the slab thickness except
at very long wavelengths ($k\to 0$) where decreasing the slab thickness
reduces the universal slope that is dictated by the presence of bulk and
surface collective oscillations.

Finally, we have found that the TPSS wavevector analysis of the correlation
surface energy, as obtained from an accurate (beyond RPA) non-oscillatory
parametrization of the xc hole density of a uniform electron gas, provides
both the exact short-$k$ limit, where LDA fails badly, and the exact large-$k$
limit, where RPA is wrong.
Hence, our calculations support the conclusion that the TPSS meta-GGA xc
density functional accurately describes the jellium surface, including not only
short-range but also long-range effects.

\acknowledgments
J.M.P. acknowledges partial support by the University of the Basque Country,
the
Basque Unibertsitate eta Ikerketa Saila, the Spanish Ministerio de Educaci\'on
y Ciencia, and the EC 6th framework Network of Excellence NANOQUANTA (Grant No.
NMP4-CT-2004-500198). L.A.C. and J.P.P. acknowledge the support of the U.S.
National Science Foundation under grant DMR-0501588.

\appendix

\section{}\label{ap2} 

Here we give an explicit expression for the wavevector-dependent contribution
$\gamma_{xc}(k)$ to the xc surface energy $\sigma_{xc}$ of a jellium slab of
background density $\bar n$ and thickness $a$, in terms of the single-particle
energies $\varepsilon_l$ and the Fourier coefficients $b_{ls}$ and
$\chi_{mn}^\lambda$ of the single-particle wave functions $\phi_l(z)$ and the
density-response function $\chi_\lambda(k_\parallel;z,z')$,
respectively.\cite{eguiluz} From Eqs.~(\ref{gamma}), (\ref{wvj6}),
(\ref{wvj5}), and
(\ref{wvj7}) and performing the integrals over the coordinates $z$ and $z'$
analytically, we find:    
\begin{eqnarray}
\gamma_{xc}(k)&=&{k_F\over\pi}\left[\int^{k}_{0}
{dk_{z}\over k}\sum^{\infty}_{m=0}\sum^{\infty}_{n=0}\alpha_{mn}
(k_z)\beta_{mn}(k_\parallel)\right.\cr\cr
&-&\left.\bar n\,a\,\bar{n}_{xc}^\mathrm{unif}(k)\right],
\label{wvj14}
\end{eqnarray}
where
\begin{equation}
\alpha_{mn}(k_{z})=2k^{2}_{z}
\frac{1-(-1)^m\cos(k_zd)}
{[k^2_{z}-(m\pi/d)^2][k^2_z-(n\pi/d)^2]}
\end{equation}
and\cite{PE2}
\begin{eqnarray}
\beta_{mn}(k_{\parallel})&=&
-\frac{1}{\pi}\int_0^1 d\lambda\int_0^{\infty}
d\omega\chi^{\lambda}_{mn}(k_{\parallel},i\omega)\cr\cr
&-&\frac{\mu_{m}\mu_{n}}{\pi d^{2}}
\sum^{l_{M}}_{l=1}(E_{F}-\epsilon_{l})
\sum^{\infty}_{l'=1}G^{m}_{ll'}G^{n}_{ll'},
\label{wvj15}
\end{eqnarray}
with $d=a+2z_0$,
\begin{equation}
\mu_m=\cases{1,&for $m=0$,\cr 2,&for $m\geq 1$,\cr}
\end{equation}
and
\begin{equation}\label{last}
G^{m}_{ll'}={1\over 2}\sum_{s=1}^\infty\sum_{s'=1}^\infty b_{ls}b_{l's'}
(\delta_{m,s-s'}+\delta_{m,s'-s}-\delta_{m,s+s'}).
\end{equation}


\begin{thebibliography}{[Vo]}
\bibitem{notem1}
We are considering here a plane-bounded semi-infinite electron system. In the
case of a jellium slab, the right-hand side of Eq.~(\ref{gamma}) should be
divided by a factor of 2.
\bibitem{notex}
For the exchange contribution to $\bar n_{xc}^\mathrm{unif}(u)$ we have used
the
non-oscillatory parametrization reported in Ref.~\onlinecite{a113}. For the
RPA correlation contribution to $\bar n_{xc}^\mathrm{unif}(u)$ we have used the
non-oscillatory parametrization reported in Ref.~\onlinecite{YPK} 
within the framework of Ref. ~\onlinecite{pw2}.
\bibitem{note1}
$z_0$ is chosen sufficiently large for the physical results to be insensitive
to the precise value employed.
\bibitem{note2} TPSS-RPA xc surface energies of a semi-infinite jellium are
indeed very close to (and usually slightly lower than) their exact RPA
counterparts, as shown in Table I of Ref.~\onlinecite{CPT}.
\end{thebibliography}
\end{document}